\documentclass[12pt]{article}
\usepackage[koi8-r]{inputenc}
\usepackage{natbib}
\usepackage[english]{babel}
\usepackage{txfonts}
\usepackage{graphicx}
\usepackage[bf,small,center]{titlesec}
\usepackage{indentfirst}
\usepackage[small]{caption2}
\usepackage{lscape}
\usepackage{alltt}

\newcommand{\nc}{\newcommand}
\def\apgt{\ {\raise-.5ex\hbox{$\buildrel>\over\sim$}}\ }
\def\aplt{\ {\raise-.5ex\hbox{$\buildrel<\over\sim$}}\ }

\newcommand{\rs}{\mbox {$R_{\odot}$}}

\newcommand{\ms}{\mbox {$M_{\odot}$}}

\newcommand{\md}{\mbox {$\dot{M}$}}

\newcommand{\myr}{\mbox {~${\rm M_{\odot}~yr^{-1}}$}}
\newcommand{\porb}{\mbox {$P_{\rm orb}$}}

\newcommand{\mhe}{\mbox {$M_{\rm He}$}}
\newcommand{\mwd}{\mbox {$M_{\rm wd}$}}
\newcommand{\tkh}{\mbox {$\tau_{\rm KH}$}}
\newcommand{\tgw}{\mbox {$\tau_{\rm GW}$}}
\newcommand{\tm}{\mbox {$\tau_{\rm m}$}}

\def\m{^m\kern-7pt .\kern+3.5pt}
\def\y{^y\kern-1.9mm .\kern+.3mm}
\def\s{^s\kern-1.7mm .\kern+.3mm}
\def\hup{^{h}\kern-2.1mm .\kern+.6mm}
\def\d{^{d}\kern-2.1mm .\kern+.6mm}
\def\P{Paczy\'{n}ski}

\def\apj{Astrophys. J. }
\def\apjl{Astrophys. J. }

\def\aap{Astron. Astrophys. }
\def\aj{Astron. J. }

\def\mnras{MNRAS }
\def\pasp{Publ. Astron. Soc. Pacific }
\def\acta{Acta Astron. }
\def\sna{SN Ia}
 
\def\am{AM~CVn}

\nc{\EK}{E_\mathrm{K}}
\nc{\K}{\:\mathrm{K}}
\nc{\Lbol}{L_\mathtarm{bol}}
\nc{\Mzams}{M_\mathrm{ZAMS}}
\nc{\Teff}{T_\mathrm{eff}}
\nc{\kms}{\:\mbox{км/с}}
\nc{\mbol}{m_\mathrm{bol}}
\nc{\req}{r_\mathrm{eq}}
\nc{\sut}{\:\mbox{сут}}
\nc{\Umax}{U_\mathrm{max}}
\nc{\vesc}{v_\mathrm{esc}}
\begin{document}

\begin{center}
\textbf{EVOLUTION OF LOW-MASS HELIUM STARS  \\[8pt]
        IN SEMIDETACHED BINARIES \\[32pt]}

L. R. YUNGELSON$ ^*$

\textit{Institute of Astronomy of Russian Academy of Sciences, Moscow} \\[6pt]

Received \today
\end{center}

\vspace{2mm}
\noindent
We present results of a systematic investigation of the evolution of low-mass (0.35\,\ms, 0.40\,\ms\ and 0.65\,\ms) 
helium donors in semidetached binaries with accretors -- white dwarfs. 
In the initial models of evolutionary sequences abundance of helium in 
the center  $0.1 \aplt  Y_c \leq  0.98 $. 
Results of computations may be applied to the study of the origin and evolutionary state of \am\  stars.
It is shown that the minimum orbital periods of the systems only weakly 
depend on the total mass of the system and evolutionary
state of the donor at RLOF and are equal to 9-11 min. 
The scatter in the mass-exchange rates at given \porb\ in the range  
$  \porb_{,min} < \porb \aplt$~40 min. does not exceed $\sim2.5$.
At $  \porb \apgt $ 20~min mass-losing stars are weakly degenerate homogeneous cooling objects and abundances  of   
 He, C, N, O, Ne in  the matter lost by them depends on the extent of He-depletion  at RLOF. 
For the systems which are currently considered as the most probable model candidates for \am\ stars with helium donors these abundances are 
$Y \apgt 0.4$, 
$  X_{\rm C} \aplt 0.3$,
$  X_{\rm O} \aplt 0.25$,
 $  X_{\rm N} \aplt 0.5\times 10^{-2}$.
At  $\porb \apgt $ 40~min. the timescale of mass-loss begins to exceed thermal time-scale 
of the donors, the latter begin to contract,  they become more degenerate and, apparently, ``white-dwarf'' and ``helium-star'' 
populations of \am\ stars merge. 

\vskip 10pt


\\
\vfill
\noindent
{$ ^*$}Email: lry@inasan.ru

\section{Introduction}
\label{sec:intro}
 
Non-degenerate helium stars in close binaries (CB) form in  so-called ``case B'' of mass-exchange,
when stars with mass  $ M \apgt (2.3  - 2.5)$\,\ms\ overflow their Roche lobes in the hydrogen-shell 
burning stage (Kippenhahn and Weigert, 1967; Paczy\'{n}ski, 1967b).  Masses of helium stars 
are $ \gtrsim 0.32$\,\ms\ (Iben, 1990; Han et al., 2002).

For the  current notions on stellar evolution, importance of low-mass helium stars
is defined by the possibility of  formation of pairs containing 
white dwarfs or neutron stars accompanied by helium stars in the course of evolution of CB. 
Orbital periods of such systems may be so short,  that angular 
momentum loss (AML) via gravitational wave radiation (GWR) enables Roche-lobe overflow (RLOF)
by helium star before helium exhaustion in the core of the latter and, under proper conditions, 
 stable mass-transfer is possible 
(Savonije et al., 1986; Iben and Tutukov, 1987; see also Yungelson, 2005a and Postnov and Yungelson, 2006).

Semidetached pairs of neutron stars with helium-star companions were not observed as yet,
though, the first computation of the evolution of a CB with a non-degenerate helium donor
was performed just for such a pair (Savonije et al. 1986). On the other hand, it is assumed that in the case when
in a semidetached binary helium star is accompanied by a white dwarf,
  such a system may be observed as an
ultra-short period cataclysmic variable 
of the \am\ type (Savonije et al., 1986).
Other hypothetical scenarios for formation of \am\ stars assume stable mass-exchange between 
white dwarfs (\P, 1967a) or mass-loss by a
 remnant of low-mass  ($\lesssim 1.5$\,\ms)
main-sequence star which overflowed its Roche lobe after exhaustion of a significant
fraction of the hydrogen in its core     ($ X_{c} \lesssim 0.4$) or even  immediately after formation of a low-mass
($ \sim0.01$\,\ms) helium core (Tutukov et al., 1985; Podsiadlowski et al., 2003)\footnote{For realization of this scenario
angular momentum loss by magnetically coupled stellar wind is also necessary.}.

AM CVn stars are important for the physics in general, since it is expected that, thanks to their extremely short
orbital periods, they, along to the close detached short-period white dwarf pairs, will be among the first 
objects which will be able to detect space-born gravitational wave antenna \textit{LISA} (Evans et. al., 1987; Nelemans et al., 2004).
More, since the distance to some of the  \am\ stars is known with sufficient accuracy, they could be used 
as verification binaries for     \textit{LISA} (Stroeer and Vecchio, 2006).

\am\ stars, despite small number of identified and candidate objects  ($\sim20 $, see the list with parameters of the systems
and references at
\begin{alltt}
\small{
www.astro.ru.nl/~nelemans/dokuwiki/doku.php?id=verification_binaries:am_cvn_stars}),
\end{alltt}
are of great interest  not only because they are potential sources of 
detectable gravitational waves, but also, for instance, because 
 their formation involves evolution in common envelopes, their
accretion disks may consist of helium or helium-carbon-oxygen mixture.
\am\ stars are potential progenitors of SN~Ia (Tutukov and Yungelson, 1979; see
also Solheim and Yungelson, 2005),
 of still hypothetical SN~.Ia 
(Bildsten et al., 2007), and of other explosive phenomena associated with
accumulation of helium at the surface of 
white dwarfs (see, e.g., Tutukov and Yungelson, 1981; Iben and Tutukov, 1991). 
Note also, that it is expected that $\sim 100$ most tight Galactic 
\am\ stars might be observed both in gravitational waves and in 
the electromagnetic spectrum (Nelemans et al., 2004).

It is still unclear which of the above-mentioned scenarios for formation of \am\ stars acts in the Nature. 
All scenarios for formation of \am\ stars imply that their progenitors pass through one or two common-envelope
stages, but since clear understanding of the processes occurring in the latter is absent, the efficiency of
common envelopes ejection remains a crucial parameter of scenarios that defines final separation of
components and, hence, possibility of formation of a 
semidetached system. Also, conditions for stable mass transfer after RLOF by a white dwarf are not
clear (see, e.g., Marsh et al., 2004; Gokhale et al., 2007; Motl et al., 2007). On the other hand, for instance,
if  \am\ stars descend from strongly evolved hydrogen-rich cataclysmic variables, only a minor fraction of them 
evolves to   $ \porb <25 $ min. which are typical for a considerable number of the \am\ stars; more, hydrogen which has to be 
present in 
the spectra of accretion disks of most of such systems is not observed as yet. 
Nevertheless, it is possible that all formation channels contribute to the population 
 of \am\ stars. Resolution of the problem of
the origin of \am\ stars is also important because observational estimates of their Galactic population are by an order of
magnitude below model estimates based on the above-described models (Roelofs et al., 2007a,b). This may point to certain
flaws in the understanding of evolution of close binaries. Note, however, that the  ``deficit'' of observed \am\   stars may be
due to numerous selection effects (see detailed discussion in Roelofs et al., 2007a,c and Anderson et al. 2005, 2008).

Nelemans and Tout (2003) noticed that the chemistry of accretion disk in an \am\ type star may serve as an identifier of
its  origin. If a helium white dwarf serves as the donor, the disk has to contain He and other H-burning products.
In the case of helium-star donor, the disk, can, along to He, contain CNO-cycle and He-burning products.  If the donor is an evolved
main-sequence star, the disk may contain H and H-burning products. In the first case abundances depend on the 
mass of the white dwarf progenitor, while in other cases it depends also on how far was the star evolved prior to RLOF.   

Evolution of CB with low-mass He-donors was computed by Savonije et. al. (1986), Tutukov and Fedorova (1989), 
Ergma and Fedorova (1990). These papers were focused, mainly,  on mass-transfer rates, ranges of orbital periods, since 
results of computations were applied to X-ray systems and SN~Ia progenitors. In the present paper we carry out a systematic 
investigation of the evolution of helium stars in semidetached binaries as a function of their mass, evolutionary state at RLOF, and 
the parameters of a binary that contains helium star. Special attention is paid to the chemical composition of the matter lost by
helium stars. In the forthcoming study (Nelemans and Yungelson, in prep.) these results will be applied for the analysis of the origin of observed \am\ stars.  

\section{Method of computations}
\label{sec:model} 
 
For our computations  we have used a specially adapted version of P.P. Eggleton evolutionary code
(1971; private comm. 2006). Equation of state, opacity tables,  and other input data are
described by Pols et al. (1995). Information on the latest modifications of the code may be found at 
\verb=www.ast.cam.ac.uk/stars/=.
In the code, abundances of $^1$H, $^4$He, $^{12}$C, $^{14}$N, $^{16}$O, $^{20}$Ne, $^{24}$Mg are computed.
The nuclear reactions  network is given in Table 1 of Pols et al. (1995). Nuclear reactions rates are taken after Caughlan and Fowler (1988),
with exception of 
$^{12}{\rm C}(\alpha,\gamma)^{16}{\rm O}$
reaction for which the data of Caughlan et al. (1985) are used. Homogeneous helium models have mass fractions of helium 
 Y=0.98, 
carbon $ X_{\rm C}=$0.00019, 
nitrogen $ X_{\rm N}=$0.01315,
oxygen $ X_{\rm O}=$ 0.00072,  
neon $X_{\rm Ne}=$0.00185, 
magnesium $ X_{\rm Mg}=$0.00068.

It was assumed that mass-exchange is conservative. Angular momentum loss via 
GWR was taken into account using standard Landau and Lifshitz (1971) formula 
 \begin{equation}
\left ( \frac {\dot{J}}{J} \right )_{\rm GWR}  = -\frac{32}{5} \, \frac{G^3}{c^5} \frac{M_{\rm He}\, M_{\rm wd} (M_{\rm He} + M_{\rm wd})}{a^{4}}.
\label{eq:gwr} 
\end{equation}
Here 
 \mhe\ and \mwd\  are the masses of components,  
$a$ is their separation. 

Mass loss rate $ \dot{M}_{\rm He}$ is related to $ \dot{J}/J $ as
\begin{equation}
\label{eq:mdot} 
\frac{\dot{M}_{\rm He}}{M_{\rm He}} = 
\left ( \frac{\dot{J}}{J} \right )_{\rm GWR} \times \left [\frac{\zeta (M_{\rm He})}{2} +
\frac{5}{6} - \frac{M_{\rm He}}{M_{\rm wd}} \right]^{-1},
\end{equation}
where
  $ \zeta (M_{\rm He}) = d\ln R/d\ln M_{\rm He}$.
The term in brackets is $\sim 1$.

\section{Results}
\label{sec:results}

\subsection{Evolution of helium stars}
\label{sec:evolution}
 
\begin{table*}[!t] 

\caption[ ]{Parameters of calculated evolutionary tracks. $M_d$, $ M_a$ -- initial 
masses of donor and accretor,  $P_0$ -- initial orbital period, $t_c$ -- time to RLOF ,  
$P_c$ -- orbital period at RLOF,  $Y_c$ -- central abundance of He at RLOF,  $t_f$ -- the age 
of the last computed model, $P_f$-- final orbital period, $M_{df}$ --  mass of the last 
computed model of the donor, $Y_{sf}$  --surface He-abundance of the last  computed model.}
\begin{tabular}{r|c|c|c|c|c|c|c|c|c|c}
\hline
No.        & $M_d,$& $ M_a,$ & $P_0$,  & $t_c$,  & $P_c$,  & $Y_c$ 
& $t_f$,  & $P_f$, & $M_{df}$ & $Y_{sf}$ \\
 & \ms & \ms & min. & $10^6$\,yr & min. & & $10^6$\,yr & min. & \ms \\
\hline
  1& 0.35 &0.5  &20   & 1.29 & 15.96 & 0.977 &427.00 & 42.06 & 0.027 & 0.976  \\ 
   2& 0.35 &0.5  &40   & 15.99 & 16.24 & 0.936 & 400.99 & 41.52 & 0.028 & 0.935 	 
    \\   
3&0.35  &0.5  &60   & 50.80 & 17.02 & 0.871 & 426.05 & 41.18 & 0.028 & 0.870 	 	 
\\    
4&0.35   &0.5  &80  & 110.71& 18.14& 0.774 & 492.51 & 40.81 & 0.027 &0.723 		 
\\  
  5&0.35  &0.5  &100 & 202.32 & 19.71& 0.642 & 615.18 & 40.35 & 0.025 & 0.640 	   
  \\
     6&0.35  &0.5  &120 & 332.36 & 22.46& 0.435 & 689.67 & 38.30 & 0.025 &0.428 	 
     \\   
7&0.35  &0.5  &140
 & 502.68 & 34.76& 0.178 & 874.27 & 37.05 & 0.022 & 0.166 	  \\ 
   8& 0.35 &0.5  &144  & 557.78 & 35.00 & 0.118 & 840.34 & 35.31 & 0. 024 & 0.104 	  
   \\
\hline   
   9&0.40  &0.6  &20   & 0.14& 19.51& 0.979& 348.53& 41.90& 0.027& 0.978 	 \\
 10&0.40  &0.6  &40   & 11.68& 19.94& 0.923& 338.05& 41.50 & 0.028 &0.920 	 \\
11&0.40  &0.6  &60   & 50.78 & 17.02 & 0.871 & 426.05 & 41.18 & 0.028 & 0.825 	  \\
12&0.40  &0.6  &80   & 84.52 & 22.42 & 0.704 & 476.41 & 41.63& 0.025 & 0.698	\\
13&0.40  &0.6  &100   &154.47 & 25.57 & 0.509 &733.73 & 41.96 & 0.020 &  0.495 	 \\
14&0.40  &0.6  &120   & 252.55 & 29.97 & 0.228 & 652.76 & 38.52 & 0.021 & 0.194	 
\\
15&0.40  &0.6  &130   & 315.30 & 29.71 & 0.066 & 528.00 & 34.08 & 0.026 & 0.035 	 
\\
\hline
16&0.40  &0.8  &20   & 0.07 & 19.71 & 0.976 & 323.79 & 42.55 & 0.027 & 0.977		  \\ 
 17&0.40  &0.8  &40  & 9.14 & 20.61 & 0.933 & 330.17 & 42.61 & 0.027 & 0.923 	  \\
18&0.40  &0.8  &60   & 30.21 & 21.67 & 0.854 &358.96 & 42.51 & 0.027 & 0.849	  \\
19&0.40  &0.8  &80   & 66.95 & 22.85 & 0.751 &431.17 & 42.60 & 0.025 & 0.744 	  \\
20&0.40  &0.8  &100  & 122.98 & 25.40 & 0.601 & 198.19 & 30.51 & 0.043 & 0.587 	\\
21&0.40  &0.8 &120   & 201.61& 28.65 & 0.376 & 568.45 & 40.12 & 0.023 & 0.353 	  \\
22&0.40  &0.8  &140   & 306.39 & 30.63 & 0.090 & 551.02 & 36.14 & 0.024 & 0.057 	 
\\
\hline
23&0.65  &0.8  &35   & 0.17 & 34.54 & 0.976 & 324.85 & 43.96 & 0.027 & 0.856 	  \\ 
24&0.65  &0.8  &40   & 2.06 & 35.29 & 0.928 & 340.97 & 44.04 & 0.026 & 0.806 	  \\
25&0.65  &0.8  &60   & 14.76 & 38.74 &0.708 & 443.33 & 43.96 & 0. 022 & 0.547 	  \\
26&0.65  &0.8  &80   & 36.29 & 44.87 & 0.396 & 353.64 & 39.99 & 0. 022 & 0.129 	 \\
27&0.65  &0.8  &85  &  42.83 & 47.07 & 0.286 & 324.29 & 38.23 & 0.023  & 0.0135 	  
\\
28&0.65  &0.8  &90   & 50.86 & 48.56 & 0.186 & 69.36 & 4.31 & 0.378 & 0.00 	 \\
 \hline
\end{tabular}
\label{tab:tracks}
\end{table*}

\begin{figure*}[t!] 
\centerline{\includegraphics[scale=0.6]{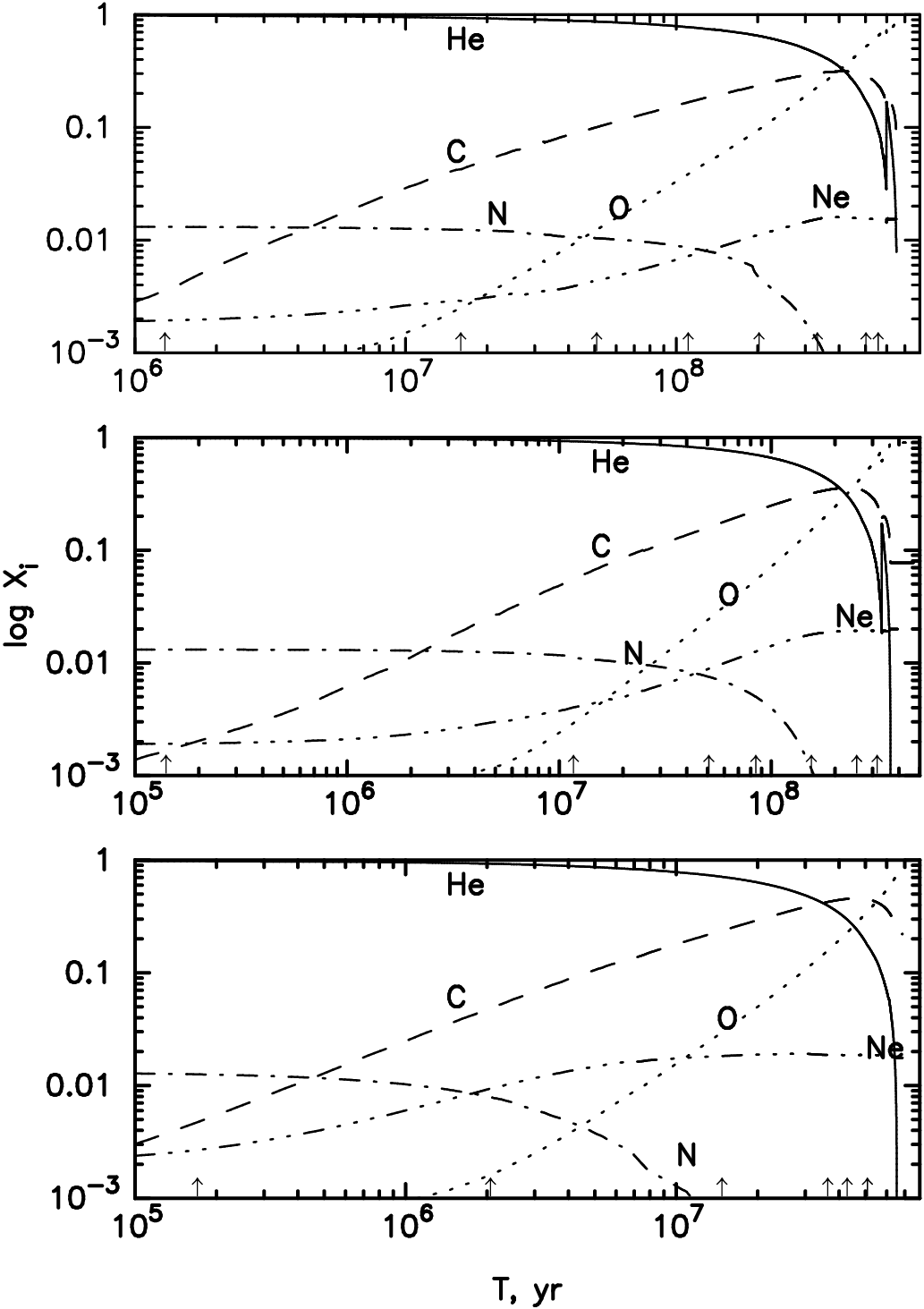}} 
\caption{ Dependence on time of abundances of He, C, N, O, and Ne in the center of 0.35\,\ms,  0.40\,\ms\ and  0.65\,\ms\  stars (top to bottom). 
 The arrows mark the epochs of RLOF by stellar models computed in the paper (Table \ref{tab:tracks}). }
\label{fig:t_comp}
\end{figure*}

Following  the model of the Galactic population of \am\ stars (Nelemans et al., 2001), 
 we have considered as typical progenitors 
of helium-donor \am\  stars the systems with masses 
$ M_{\rm He}+M_{\rm wd} $ =   (0.35 + 0.5)\,\ms, (0.4 + 0.6)\,\ms, and (0.4 + 0.8)\,\ms.
In addition, we have considered a  pair  (0.65+0.8)\,\ms. If helium-accreting white dwarfs avoid double- (or ``edge-lit-'') detonation which may destroy the dwarf, such a system may belong  to progenitors of 
\am\ stars, but not to  the most ``fertile'' of them.  
As the second parameter of computations we considered initial orbital period of the system, i.e., the period of the system
immediately after completion of the common envelope stage that resulted in  
formation of a helium star. The range of $\porb_{,0}$\ for every system was chosen in such a way that the set of
computed models included both stars that filled their Roche lobe virtually unevolved
and stars that had at the instant of RLOF radii close to the maximum of the radii of low-mass helium stars which are attained 
at $ Y_c \simeq 0.1 $  (see Table.~\ref{tab:tracks} and Fig.~\ref{fig:t_comp}). The maximum period in the set of
initial orbital periods for every computed $ \mhe + \mwd $ pair is, in fact, the limiting initial period which still  allows 
RLOF and formation of an \am\ star. 
Nuclear evolution of He-stars with mass  $M \lesssim 0.8 $\,\ms\ terminates after helium burning stage and they evolve directly 
into white dwarfs (\P, 1971).

The lifetime of helium stars is short. According to our computations for  (0.35 - 0.65)\,\ms\ mass range 
\begin{equation}
\label{eq:age}
 t_{\mathrm He} \approx 10^{6.95}\,M_{\rm He}^{-4.1}(1+M_{\mathrm He}^{3.74}), 
\end{equation}
where time is in yr, masses in \ms. 
During He-burning, the radii of helium stars increase by $\sim 30\% $ only. Therefore, the crucial factor which defines
the range of the orbital periods of the precursors of \am\ stars is AML. 
In the Nelemans et. al. (2001) model,  in $ \simeq 90\% $
of the systems helium stars overflow their Roche lobes during first
  $ \simeq 50\%$ of their lifetime. Figure~\ref{fig:t_comp}
shows that this time-span corresponds to the reduction of  $Y_c $ to
$\simeq 0.5 $. Thus, most of the \am\ systems might have initial orbital periods up to 100 -- 120 min.

Evolution of mass-losing stars is followed to $M \simeq (0.02 -0.03)\,\ms$. Continuation of computations is
hampered by the absence of adequate EOS and low-temperature opacity tables in our code. 
However, for  the analysis of the scenarios of formation and evolution of \am\ stars which involves information on
the elemental abundances in the donor-star this is  not important, since the chemical composition
of the stars in our case becomes  ``frozen'' even before the minimum of \porb\ is reached, because of the drop
of the temperature in the core and switch-off of nuclear burning (see below).

\begin{figure*}[t!] 
\centerline{\includegraphics[scale=0.6,angle=-90]{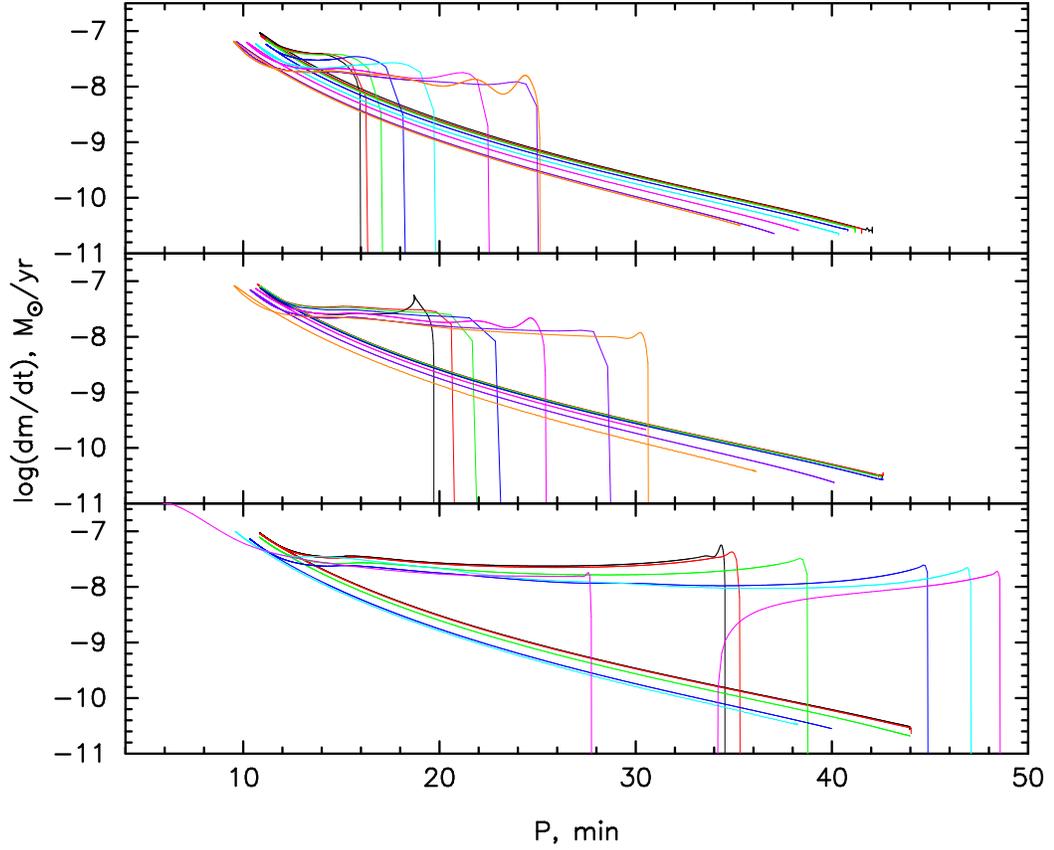}}
\caption{Dependence of mass-loss rate on the period for computed evolutionary sequences for the systems with initial mass of
components
 $0.35\,\ms\ + 0.5\,\ms$,  $0.40\,\ms\ + 0.8\,\ms$,  $0.65\,\ms\ + 0.8\,\ms$
 (top to bottom). }
\label{fig:mdot}
 \end{figure*}  

\begin{figure*}[t!] 
\begin{minipage}[t]{0.49\textwidth}
\includegraphics[scale=0.38,angle=-90]{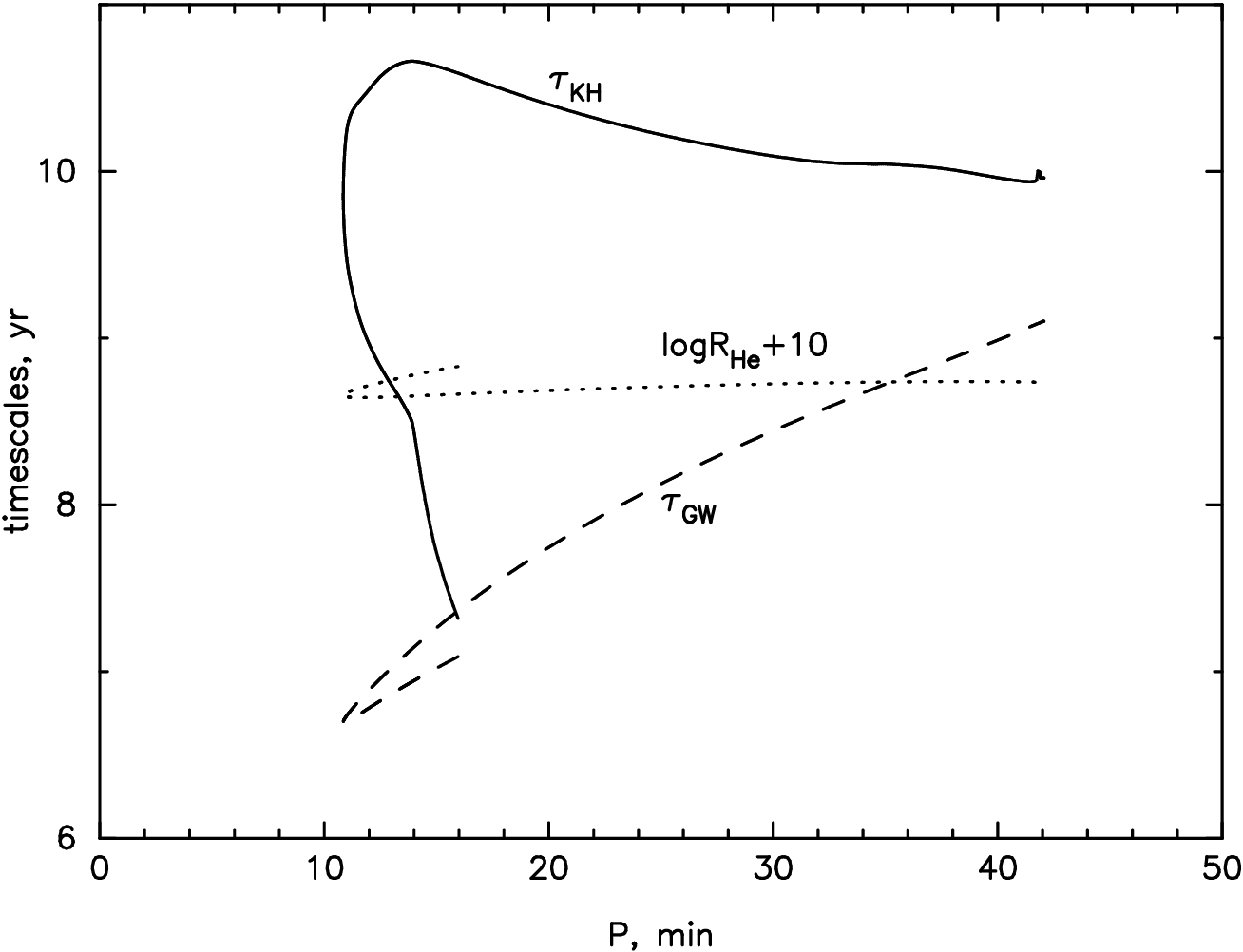}
\end{minipage}
\hfill
\begin{minipage}[t]{0.49\textwidth}
\includegraphics[scale=0.38,angle=-90]{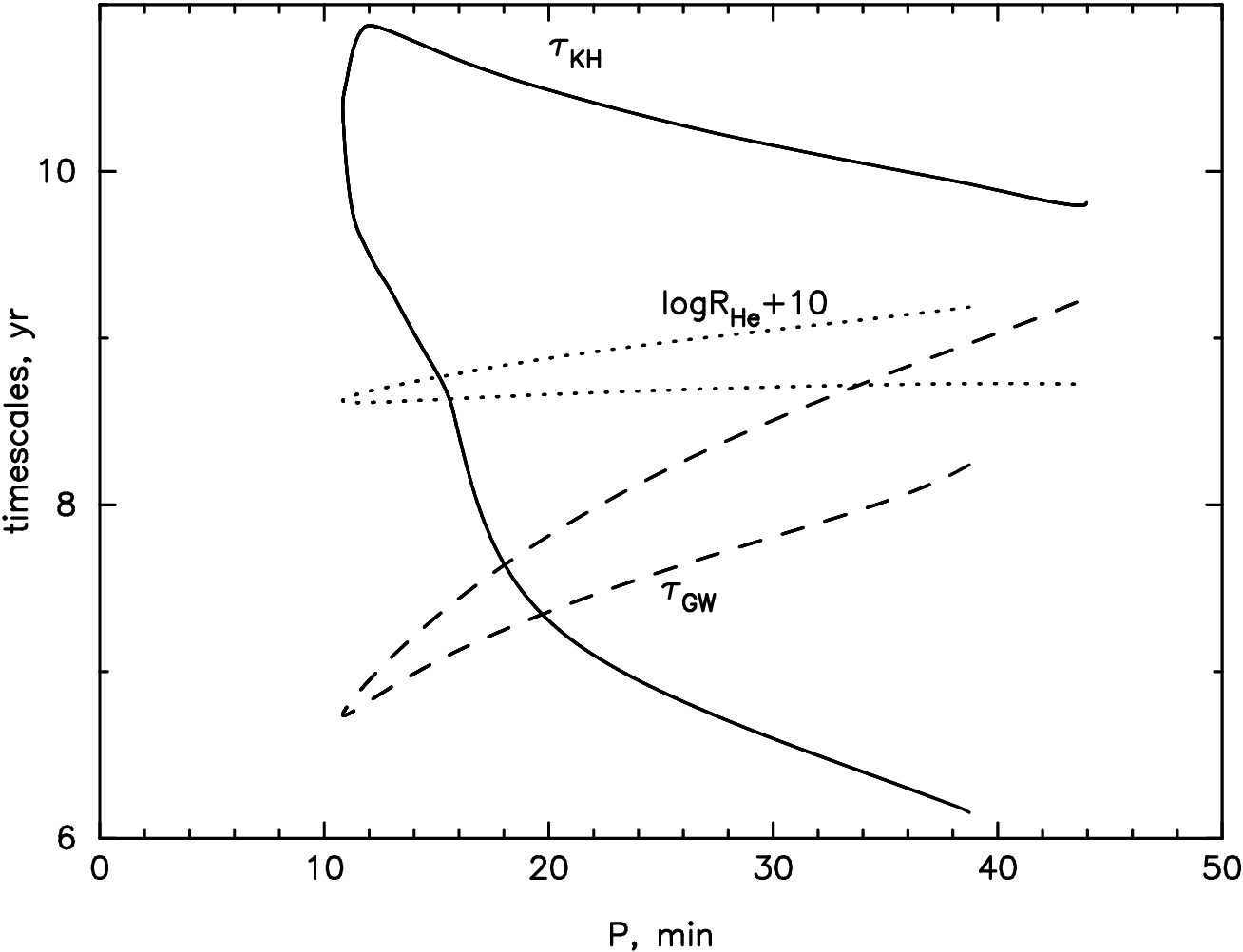}
\end{minipage}
\caption{ Variation of thermal timescale of helium star  \tkh, angular momentum loss timescale  \tgw\ and donor radius (in \rs) 
 in  the binaries with
initial parameters  $M_{2,0}=0.35$\,\ms,
$M_{1,0}=0.5$\,\ms\ and  $P_{\rm orb,0}=20$\,min. (left panel) and $M_{2,0}=0.65$\,\ms,
$M_{1,0}=0.8$\,\ms\  and  $P_{\rm orb,0}=60$\,min. (right panel). A constant is added to the radii for combination of the curves in the same plot.  }
\label{fig:tscales}
\end{figure*}

Figure \ref{fig:mdot} shows dependence of the mass-loss rate by Roche-lobes filling helium stars on \porb.
The behavior of low-mass components of CB evolving under the influence of AML via GWR, like for 
``usual'' hydrogen-rich cataclysmic variables, is defined by the relation between the timescale of stellar 
evolution $ \tau_{\rm ev} $, thermal timescale of the star $ \tau_{\rm KH} $ and AML timescale \tgw\
(Faulkner, 1971; \P, 1981; Savonije et al., 1986). The relation between the timescales itself depends on the mass of the star,
its evolutionary state at RLOF and total mass of the binary (see  
Fig.~\ref{fig:tscales}).

In the system (0.35+0.5)\,\ms, $\porb_{,0}$=20\,min. with initially 
virtually unevolved donor, $ \tau_{\rm KH} >  \tau_{\rm GW} $
after the RLOF. 
Due to AML separation of components and Roche-lobe radius decrease.  
The star which is not in thermal equilibrium  reacts to the mass loss
by an increase of mass-loss rate.
 But since mass-transfer acts in opposite direction (moves components apart),
$\dot{M} $\ does not change significantly.  A considerable fraction of 
the energy generated by nuclear burning is absorbed in the envelope of the star, resulting in decrease of luminosity and 
further growth of  $ \tau_{\rm KH}$. With decrease of the stellar mass the significance of nuclear burning is rapidly diminishing.
convective core disappears, while a surface convective zone appears; during this stage  $ \tau_{\rm KH} \gg \tau_{\rm GW} $. 
As \porb\ drops, \tgw\  continues to decrease, resulting in increase of \md. At certain instant the effect of mass-transfer starts 
to dominate, orbital period reaches the minimum, while \md\ -- the maximum.  
With increase of \porb\ the timescale of AML increases, but since the 
star becomes more degenerate and has convective envelope, the power of 
$M-R$ relation is negative 
and mass loss continues, but \md\ rapidly drops.  

If the donor is initially more massive and more evolved at the instant of 
RLOF, it retains thermal equilibrium for a longer time and \md\ is 
initially completely defined
by AML. But \tkh\ increases with decrease of \mhe, while \tgw\ decreases 
and, starting from a certain moment, 
evolution proceeds like in the lower-mass and less-evolved system described above. Evolution of characteristic timescales for the system (0.65+0.8)\,\ms, $P_{\rm orb,0}$ =60\,min. is similar to that described by Savonije et al. (1986) for a 
(0.6+1.4)\,\ms\ system with  $Y_c=0.28$ at RLOF which is usually quoted as ``typical''.

The minimum periods of  the systems with helium donors are confined to a narrow range of $\porb = (9.3 -- 10.9)$\,min,
masses of stars at $\porb_{, min}$ are 0.20\,\ms -- 0.26\,\ms. The models that overfilled their Roche lobes in the 
 advanced stages of
evolution  reach lower periods and have at minimum period lower masses
than initially less evolved donors.  This may be related to the lower abundance of He and enhanced abundance of heavy elements.

\subsection{Mass-radius relation}
\label{sec:m_r}

\begin{figure*}[t!]  
     \centerline{\includegraphics[scale=0.6,angle=-90]{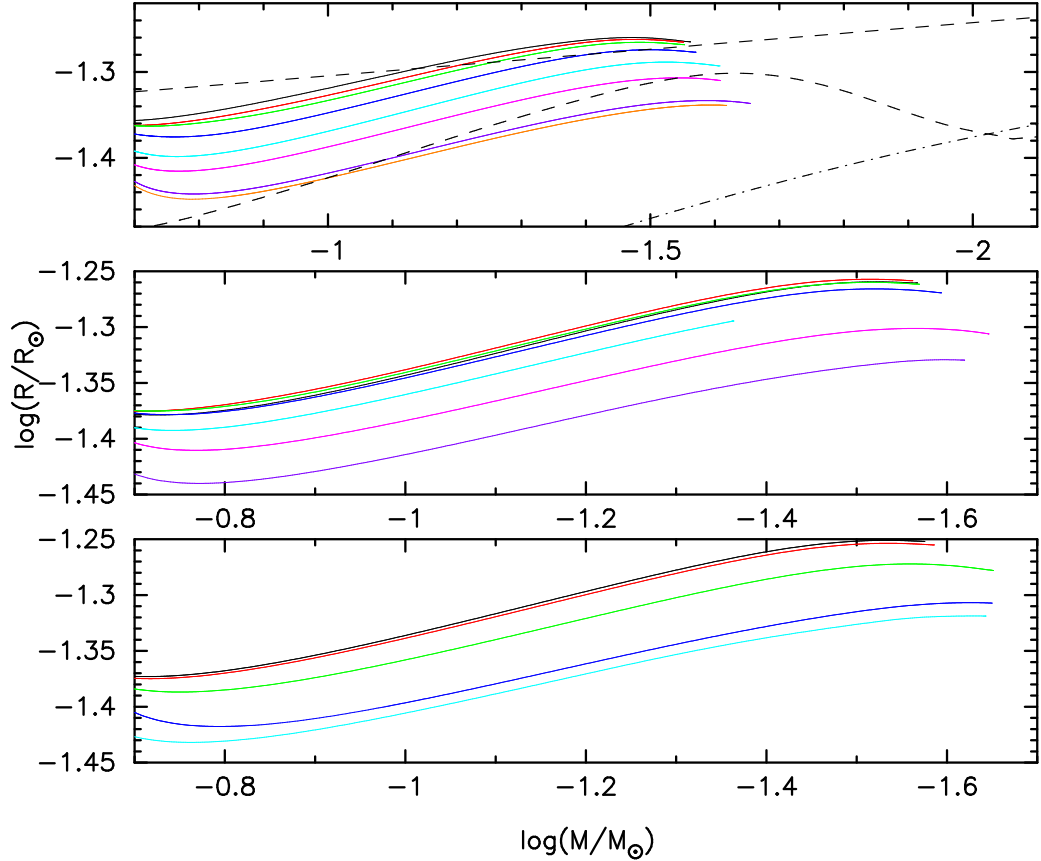}}
     \caption{Dependence of model radii on stellar mass for the systems 
 $0.35\,\ms\ + 0.5\,\ms$ (upper panel),
  $0.40\,\ms\ + 0.6\,\ms$ (middle panel),
 $0.65\,\ms\ + 0.8\,\ms$ (lower panel).
Less evolved systems are at the top of every panel.
In the upper panel upper dashed line shows 
$M-R$ relation (\ref{eq:mr_tf}), lower dashed line shows $M-R$ relation for 0.3\,\ms,
$\log \psi_0=1.1$  helium white dwarf (Deloye et al., 2007) 
and dash-dotted line shows the  fit to $M-R$ relation for zero-temperature He white dwarfs
suggested in the latter paper. }
\label{fig:mr}
\end{figure*}

\begin{figure*}[t!]    %
\centerline{\includegraphics[scale=0.5,angle=-90]{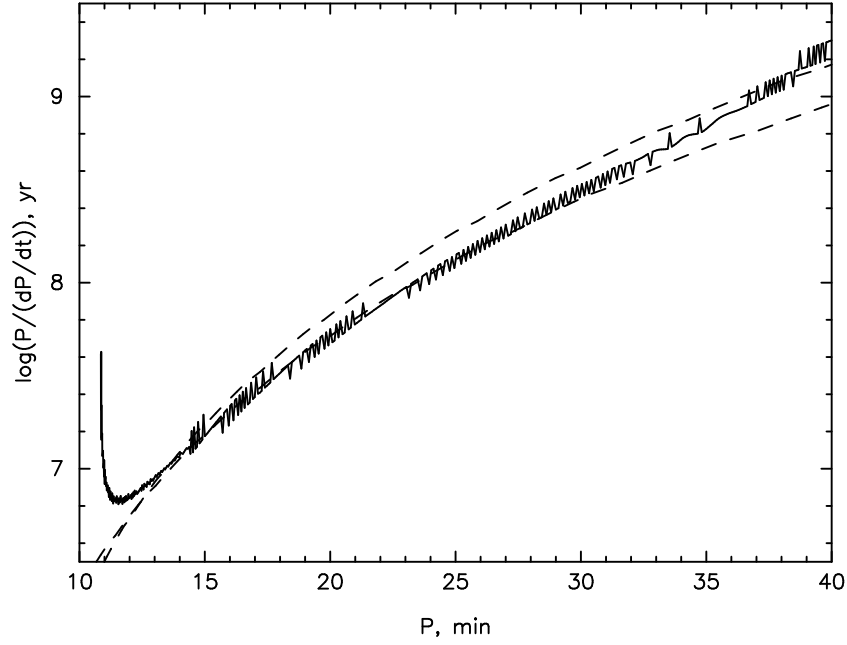}}
\caption{Dependence of $P_{\rm orb}/\dot{P}_{\rm orb}$\ on  \porb\  for the 
0.35\,\ms\ + 0.5\,\ms, $\porb_{, 0}=20$\, min. system. 
Solid line -- model calculations, upper dashed line -- dependence obtained using $M-R$ relation
(\ref{eq:mr_tf}), 
lower dashed line  -- dependence obtained by means of  relation 
(\ref{eq:mr_lry}).}
\label{fig:pdotp}
\end{figure*}

All existing computations of the evolution of initially non-degenerate helium donors  are discontinued at 
$\mhe \simeq 0.02$\,\ms. However, starting from $ \porb \simeq 20$\,min. these stars are, in fact, weakly degenerate 
homogeneous objects (the parameter of degeneracy  $\psi \approx 8.33\times 10^7 \rho_c T_c^{-3/2} \sim 10$).
This allows to consider qualitatively their further evolution using results of computations for  mass-losing arbitrary 
degenerate helium dwarfs (Deloye et al., 2007). Deloye and coauthors have shown that, when the mass of the dwarf
decreases to
$ \simeq 0.01 - 0.03$\, \ms, their \tkh\ becomes comparable to the timescale of mass loss \tm\ and further
$\tkh < \tm$. The stage of adiabatic expansion terminates, the donor gradually cools, becomes more degenerate and 
$M-R$ relation approaches relation for fully degenerate configurations. Expected picture of evolution is confirmed by
Fig.~\ref{fig:mr} which shows variation of stellar radii with decrease of mass. In addition to our data, 
we show in  Fig.~\ref{fig:mr} mass-radius relation for degenerate helium white dwarf with initial mass of 0.3\,\ms\ and
initial degeneracy parameter  $ \log \psi =1.1 $\ (the least degenerate initially model,  C. Deloye priv. comm.) and $M-R$ relation for zero-temperature He white dwarfs (Deloye  et al. 2007).
It is seen that the morphology of $M-R$ curves for helium stars and helium dwarfs is similar and, hence, one may expect
that the remnants of helium stars will gradually approach the curve for the remnants of helium dwarfs\footnote{When comparing
our results with those of Deloye et. al. one must have in mind the differences in initial masses of models and degeneracy 
parameters, as well as in the input parameters of different codes.}. Note, 
Deloye et al. in fact predicted this effect,   
based on consideration of the physics of cooling 
objects and they noticed also that some of the  
evolutionary tracks published by Tutukov and Fedorova (1989) have features
described above.

Our results, combined with results of Deloye et al. (2007) suggest that  at $\porb \apgt  40$\,min. two families of AM CVn stars  -- the 
``white dwarf'' one and the ``helium star'' one -- merge and the origin of the donors may be then identified by 
their chemical composition only (see \S\ref{sec:chem}).  

Our conclusions concerning $M-R$ relation may have certain implications for population models of \am\ stars.
Existing models (Tutukov and Yungelson, 1996; Hils and Bender 2000; Nelemans et al., 2001, 2004) were carried out under assumption that 
the power of $M-R$ relation is constant, keeps also for  $M_{\rm He} \aplt 0.02$\,\ms\ and that the evolution of stars is restricted by the Hubble time only. 
This assumption allowed to use for modeling Eq.~(\ref{eq:mdot}) with constant   $ \zeta (M_{\rm He})$.
Such an approach is, apparently, not justified and one has to use an $M-R$ relation that takes into account growing degeneracy
 of donors with mass loss.

Figure   \ref{fig:mr} shows that the $M-R$ relation  (in solar units) 
\begin{equation}
\label{eq:mr_tf}
R \approx  10^{-1.367}M_{\rm He}^{-0.062},
\end{equation}
suggested by Nelemans et al. (2001) using results of Tutukov and Fedorova (1989)  does not agree well with the results of
more modern computations and is not valid for  $ \porb \apgt 35$\,min.

In the period range $\porb \approx 10-35$\, min which hosts 10 out of 17 \am\ stars with estimated orbital periods,
the radii of the models with initial masses  (0.35-0.40)\,\ms\ may be approximated as 
\begin{equation}
\label{eq:mr_lry}
R \approx 10^{-1.478}\left(\frac{P_{\rm orb,0}}{20 {\rm min}}\right)^{-0.05} M_{\rm He}^{-0.16}\left(\frac{0.35}{M_{\rm He,0}}\right)^{0.345}
\end{equation}
where  $P_{\rm orb,0}$ and $M_{\rm He,0}$ are initial orbital period
 of the system and initial mass of the donor and $M$ and $R$ are in
solar units, \porb\ - in min. 

The time spent by a star in the certain range of orbital periods is proportional to $ \porb/\dot{P}_{\rm orb} $.
In Fig.~\ref{fig:pdotp} we compare dependencies of $ \porb/\dot{P}_{\rm orb} $ on \porb\ 
for the system (0.35+0.50)\,\ms, $\porb_{,0}=20$ obtained in
evolutionary computations and obtained by means of Eqs.~(\ref{eq:mr_tf}) and
(\ref{eq:mr_lry}). It is evident that in the  $\porb \approx (10-35)$\,min.
range Eq.~(\ref{eq:mr_tf}) results in the 20-25\% difference in the number of systems. Note, however, that this 
discrepancy is not very significant if one takes into account all uncertainties involved in the modeling of the population of
\am\ stars.

\subsection{Masses of donors in \am\ stars}
\label{sec:masses}

\begin{figure*}[t!]    %
\centerline{\includegraphics[scale=0.5,angle=-90]{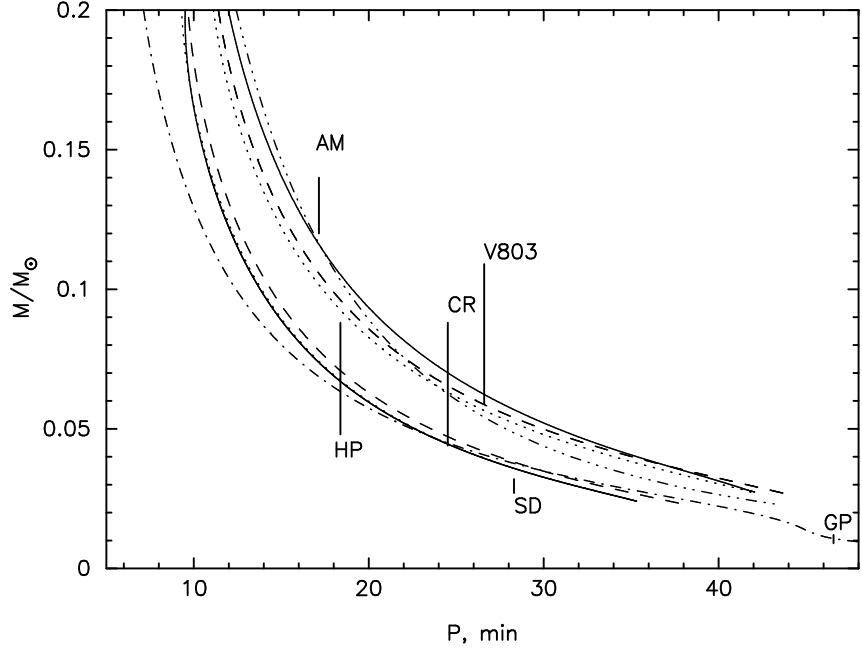}}
\caption{ Orbital period -- donor mass relations. For every given combination of initial masses of components
we show   $\porb-M$\ relation  for the system with initially least evolved and most evolved donor.
$0.35\,\ms\ + 0.5\,\ms$ -- thin solid lines,
$0.40\,\ms\ + 0.6\,\ms$ -- dashed lines,
$0.65\,\ms\ + 0.8\,\ms$ -- dotted lines.
For initially evolved donors the lines practically coincide.
Dash-dot-dot-dot line shows 
 $\porb-M$ relation for  $0.35\,\ms\ + 0.5\,\ms$, $P_{\rm orb,0}$ system based on approximation    (\ref{eq:mr_tf}).
Dot-dash line shows 
$\porb-M$ relation for $ M_0=0.3$\,\ms, $\log \psi_0=1.1$ helium white dwarf from Deloye et al. (2007).
Vertical bars show the ranges of donor-mass estimates for  
\am,  HP~Lib, CR~Boo, V803~Cen, SDSS~J0926+3624, and GP~Com  (see references in the text).}
\label{fig:pm}
\end{figure*}
Relatively accurate masses of components are derived only for  SDSS~J0926+3624
with \porb=28.3\,min, a unique partially eclipsing \am\ type system  (Marsh et al., 2007):   $ M_{\rm wd}=0.84\pm0.05$\,\ms,
 $ \mhe=0.029\pm0.002$\,\ms. 
In the $ P-M $ diagram (Fig.~\ref{fig:pm}) the donor of  SDSS~J0926+3624 is located both below the curve corresponding to the
initially  most evolved nondegenerate helium donors and the curve that describes the least degenerate white dwarfs. Apparently, this system belongs to the
latter family of stars (see also Deloye et al., 2007). Note however, that at the moment when \porb=28.3\,min, the remnants
 of initially substantially evolved donors become oxygen-neon white dwarfs (see Figs.~\ref{fig:ab035} --
\ref{fig:ab065}).   
Though existing models of the population of \am\ stars suggest that the probability of RLOF by a far-evolved helium star is low,
only analysis of chemical composition of the donor will be able to identify the scenario of the origin of SDSS~J0926+3624.

We show in Fig. \ref{fig:pm} also mass estimates for several other \am\  stars after Roelofs et al. (2006; 2007b). For \am\ the masses
of components are based on the kinematical data:   $\mwd= (0.68 \pm 0.06)$\,\ms,
$\mhe= 0.125\pm 0.012$\,\ms. For HP Lib, CR Boo, and   V803 Cen masses of the donors are derived indirectly, 
assuming that for these systems one may use the relation between superhump period excess
and mass ratio of components derived by Roelofs et al. (2006)  for \am:
$ \varepsilon(q) = 0.12q $. 
For GP Com the range of possible donor mass is based on the  limits for its luminosity and assumptions about the nature of the star. 
The estimated masses of the donors agree with assumption that they descend from nondegenerate helium stars (see also Roelofs et al., 2007a). Note however, that there are no traces of He-burning products in the spectra of these stars. 

Spectral lines of nitrogen, neon, and helium were discovered in GP Com (Lambert and Slovak, 1981; Marsh et al., 1991, 1995; 
Strohmayer, 2004). In particular, Strohmayer found the following abundances:
$Y = 0.977\pm 0.002$, 
$X_{\rm N} = 1.7 \pm 0.1 \times 10^{-2}$, 
$X_{\rm O} = 2.2 \pm 0.3\times 10^{-3}$, 
$X_{\rm Ne} = 3.7 \pm 0.2 \times 10^{-3}$,
 $X_{\rm C} < 2 \times 10^{-3}$,
 $X_{\rm Mg} < 1.6\times 10^{-4}$.
Considerable excess of nitrogen compared to oxygen and carbon points to the enrichment of stellar matter by the 
products of CNO-cycle. Such abundances may be typical for the remnants of the least massive possible progenitors
  ($\mhe_{,0} \sim 0.3$\,\ms) which overfilled Roche lobe soon after TAMS, before helium started to burn in their interiors. 
In such stars abundances in the matter lost by them is virtually the same during whole course of evolution and close to the ones
found by Strohmayer.

\begin{figure*}[ht!] 
\centerline{\includegraphics[scale=0.5]{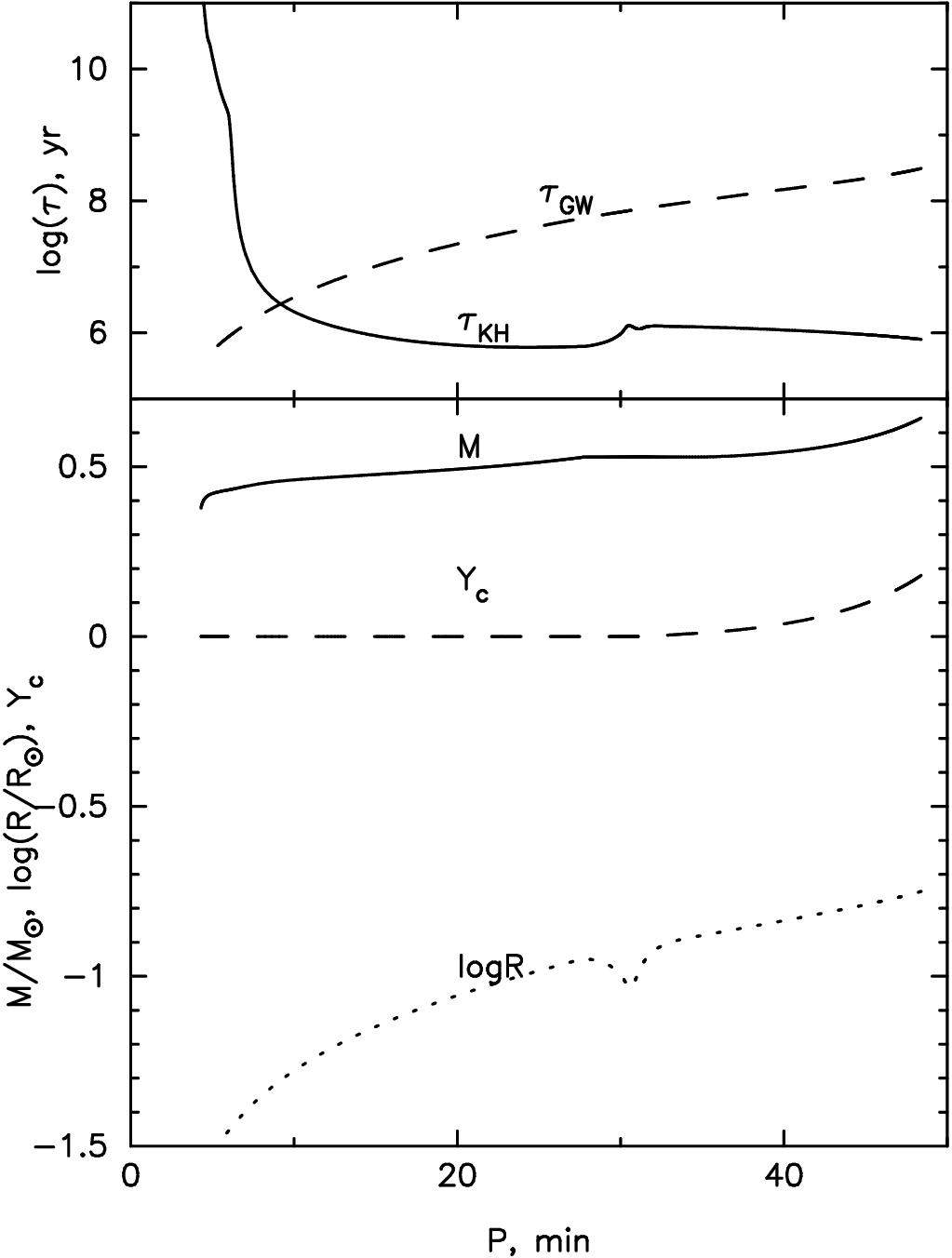}}
\caption{ Lower panel: dependence of mass, radius and central helium abundance on time for the model with initial mass  $M_{2,0}=0.65$\,\ms\ in the system with 
 $M_{1,0}=0.8$\,\ms,  $P_{\rm orb, 0}=90$\, min.
Upper panel: Variations of angular momentum loss timescale and thermal timescale of the donor. }
 \label{fig:0650890}
 \end{figure*}

\subsection{The system  (0.65+0.8)\,$\mathbf{ M_{\odot},  P_{\rm orb,0}=90}$\,min.}
\label{sec:0650890}

The system with initial parameters (0.65+0.8)\,\ms, $P_{\rm orb,0}=90$\,min. is of special interest because of its ``nonstandard''\ 
evolution which may illustrate transitional behavior between ``white dwarf'' and ``helium star'' 
scenarios of formation of \am\ stars. At the instant of RLOF abundances in the center of the donor are
   $Y \approx 0.19$, $X_{\rm C} \approx 0.45$, 
$X_{\rm 0} \approx 0.34$, nitrogen is destroyed already. At difference to lower mass stars in which nuclear burning  terminates
after loss of several hundredth of \ms, in this  more massive and hotter star it lasts longer. When the mass decreases to
  $M_2 \approx 0.52$\ms\ and  $Y_c$ becomes $\approx 0.008$, the star begins to contract. Mass-exchange interrupts for $\simeq  2.2$\, Myr\
and resumes in the helium-shell burning stage (Fig.~\ref{fig:0650890}). 
After the loss of additional  $\simeq  0.1$\,\ms\ nuclear
burning ceases and the star is now  a ``hybrid'' white dwarf with $\simeq  0.3$\,\ms\ carbon-oxygen-neon core  and $\simeq  0.1$\,\ms\ 
helium envelope with an admixture of C and O. The timescale of AML \tgw\ continues to decrease, while \tkh\ increases. 
Evolution of the donor is accompanied by its cooling  and decrease of radius. The dwarf experiences a stage of adiabatic contraction
similar to the one that is typical for the initial stages of mass-exchange in the systems with degenerate donors (Deloye et al., 
2007).  
In the last computed model of the sequence, \porb=4.3 min, \mhe=0.375\,\ms,
degeneracy parameter of the center of the star is $\psi \approx 14$. One may
expect that further evolution will be qualitatively similar to the evolution of helium white dwarf-donors: contraction phase will
be followed
by expansion which will continue until mass will decrease to  $\simeq  0.01$\,\ms\ and  thermal timescale will become comparable 
to the mas-loss timescale and a stage of contraction to the fully degenerate configuration will ensue.  The minimum \porb\ of
the system will be several minutes. Note, that at difference to He white-dwarf
donors which retain their original chemical composition during evolution,
in the case of He-star descendant under consideration, the donor will become an oxygen-carbon 
white dwarf, with an admixture of neon $(X_{\rm O} \approx 0.27$, $X_{\rm C} \approx 0.71,$
see Fig.~\ref{fig:t_comp}).

Total mass of this system is 1.45\ms. If unstable He-burning at the surface of white dwarf
does not decrease its mass below Chandrasekhar limiting mass, such a system may be  a \sna\ progenitor.

\section{Evolution of stellar chemical composition}
\label{sec:chem}

\begin{figure*}[t!]
\centerline{\includegraphics[scale=0.6,angle=-90]{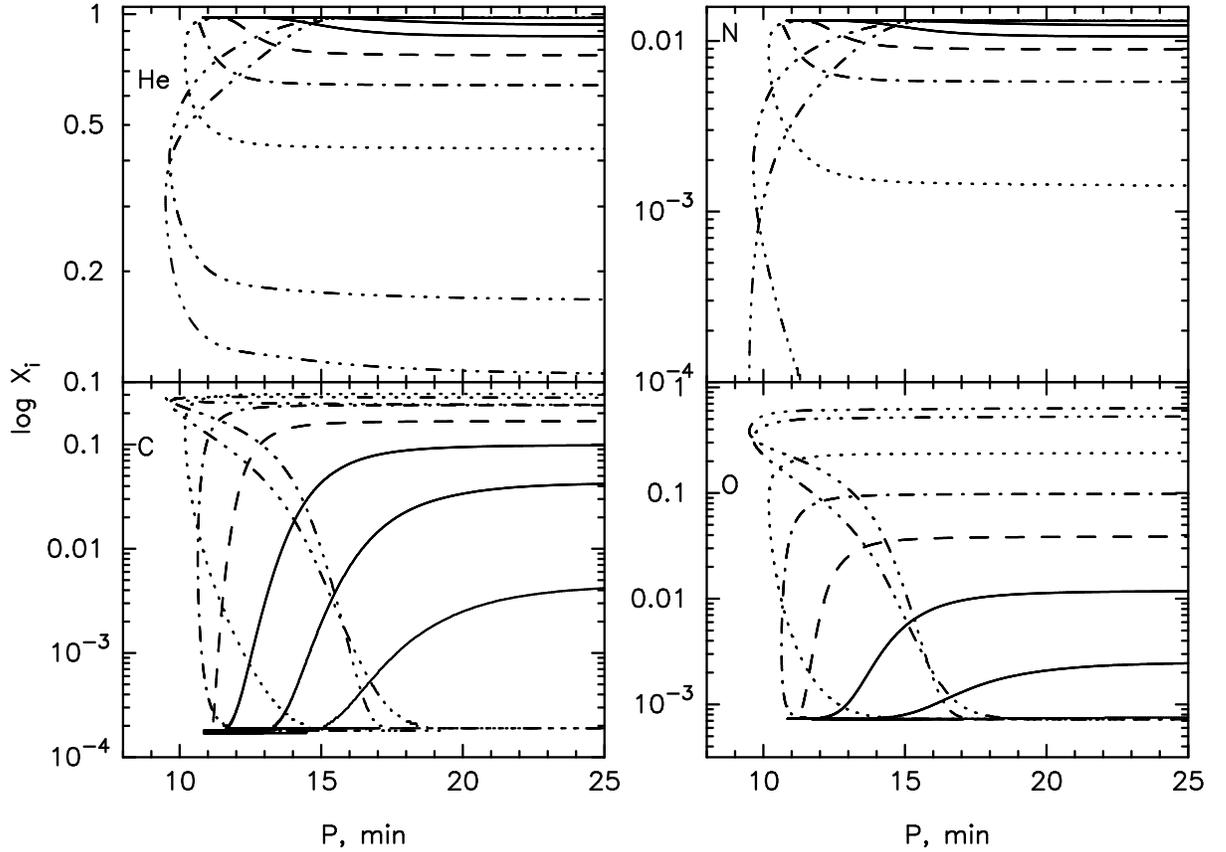}}
\caption{Variations of He, C, N, and O abundances in the accreted matter in the system with initial masses of components  
0.35\,\ms\ and 0.5\,\ms. Solid lines are for systems with initial orbital periods $P_{\rm orb, 0}=$20, 40 and 60 min.,
dashed lines  -- for $ P_{\rm orb, 0}=$80 min.,
dash-dot lines -- for $ P_{\rm orb, 0}=$100 min., 
dotted line -- for $ P_{\rm orb, 0}=$120min., 
dash-dot-dot-dot line -- for $ P_{\rm orb, 0}=$140 and 144\, min. For the system with  $ P_{\rm orb, 0}=$20\, min.
He-abundance virtually does not change. All abundances do not change any more at 
$ P_{\rm orb} \gtrsim $25\,min. 
 }
\label{fig:ab035}
\end{figure*}

 \begin{figure*}[t!]
\centerline{\includegraphics[scale=0.6,angle=-90]{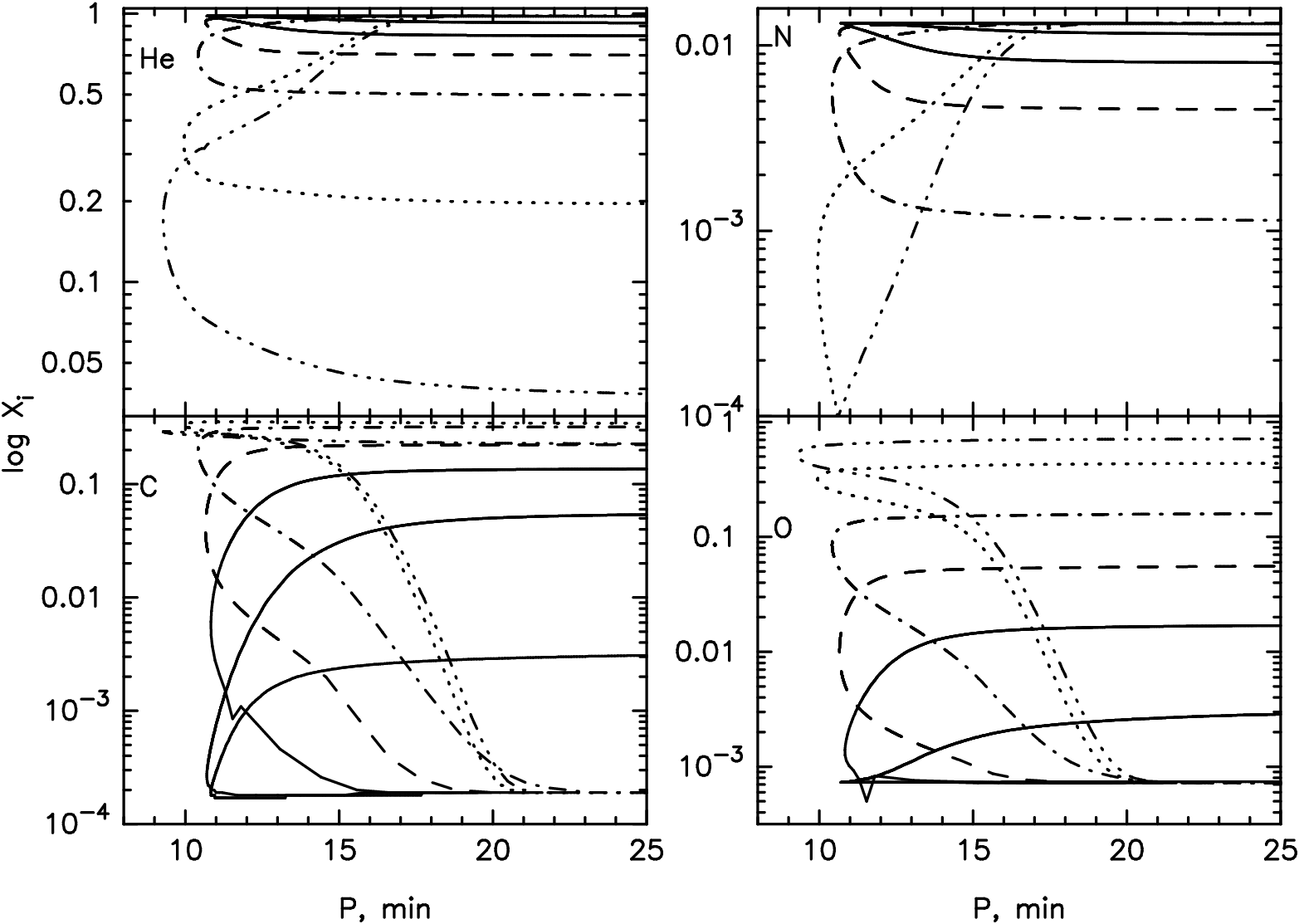}}
\caption{Same as in Fig.~\ref{fig:ab035} for the system with initial masses of components 0.4\,\ms\ and 0.6\,\ms.
Solid lines show abundances for the systems with $ P_{\rm orb, 0}=$20, 40 and  60 min.,
dashed lines   --  for  $ P_{\rm orb, 0}=$ 80 min.,
dash-dot lines   --  for  $ P_{\rm orb, 0}=$100 min., 
dashed line  -- for  $ P_{\rm orb, 0}=$120 min., 
dash-dot-dot-dot line -- for  $ P_{\rm orb, 0}=$130 min. }
\label{fig:ab04}  
\end{figure*}

\begin{figure*}[h!]
\centerline{\includegraphics[scale=0.6,angle=-90]{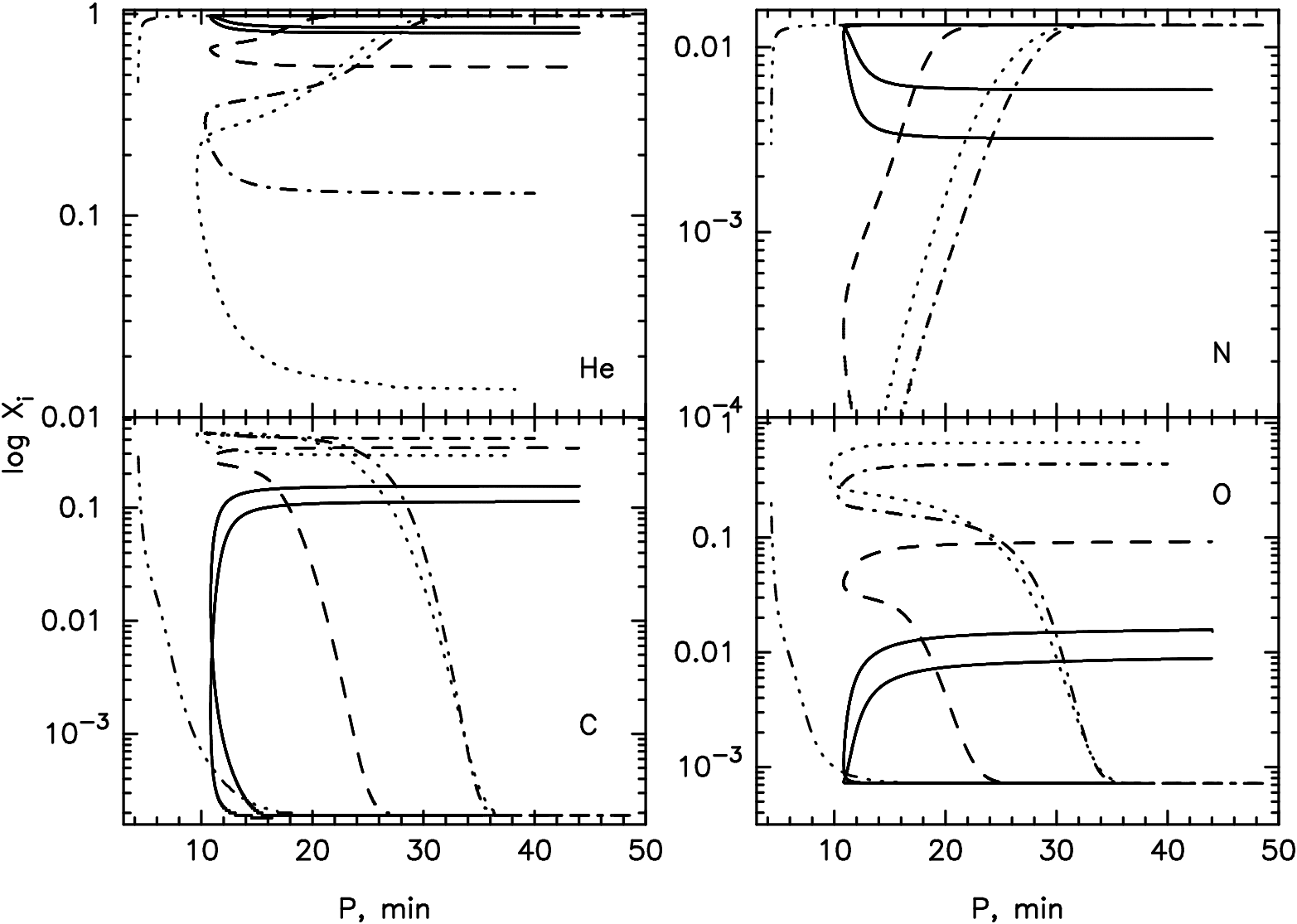}}
\caption{ Same as in Fig.~\ref{fig:ab035} for the system with initial masses of components  0.65\,\ms\ and 0.8\,\ms.
Solid lines show abundances for the systems with $ P_{\rm orb, 0}=$35 and  40 min.,
dashed lines   --  for  $ P_{\rm orb, 0}=$60 min.,
dash-dot lines  --  for  $ P_{\rm orb, 0}=$80 min., 
dashed line  -- for  $ P_{\rm orb, 0}=$85 min., 
dash-dot-dot-dot line -- for  $ P_{\rm orb, 0}=$90 min.}
 \label{fig:ab065}
\end{figure*}

Figures  \ref{fig:ab035} --  \ref{fig:ab065} show the evolution of chemical composition of matter lost by stars upon
RLOF. As we mentioned before, in the stars with initial masses 0.35 and 0.40\,\ms\ nuclear burning terminates soon after RLOF.
Simultaneously, disappears the convective core. When \porb\  is still larger than $\porb_{, min}$,  abundances in the lost matter are virtually constant and
correspond to the abundances in the matter that experienced hydrogen burning via CNO-cycle. In the period range 
from $\porb_{, min}$\ to $\porb \approx (15 - 20$)\,min. the convective core of initial model is uncovered.  The donors in initially
most wide systems are an exception, since they had large cores and chemical composition of the lost matter starts to change
already before $\porb_{,min}$. But this change happens over  such a short time that it is highly unlikely to observe a star 
in this ``transition'' state. Thus, in the range of \porb\ that harbors almost all known \am\ stars chemical composition of the matter accreted by white dwarf 
is defined by the extent of helium exhaustion at the instant of RLOF (compare Fig.~\ref{fig:t_comp} and Figs.~\ref{fig:ab035} --  \ref{fig:ab065}).
 
As we mentioned above, it is expected that the  overwhelming majority of precursors of \am\ stars in the ``helium star'' channel
of formation were formed  with $\porb <$ 100 -- 120 min. Then, typical abundances in the transferred matter are 
$Y \apgt 0.4$, 
 $  2\times 10^{-4} \aplt X_{\rm C} \aplt 0.3$,
$ 7\times 10^{-4} \aplt X_{\rm O} \aplt 0.25$,
$ 5\times 10^{-4} \aplt X_{\rm N} \aplt 0.5\times 10^{-2}$.
However, one cannot exclude that abundance of nitrogen is much lower or it is absent at all; diminished  (or zero) $X_{\rm N}$
has to correlate with relatively low $Y$ and enhanced $X_{\rm C}$ and $X_{\rm O}$.

We did not plot in Figs.~\ref{fig:ab035} --  \ref{fig:ab065} variations of the Ne abundance. It also changes rapidly 
by $\porb_{, min}$ and at  
$\porb > $20 min. neon abundance   
$2 \times 10^{-3}  \aplt X_{\rm Ne} \aplt 2 \times 10^{-2}$.
Variation of  $ X_{\rm Ne} $ is, evidently, too small to serve as identifier of the scenarios of \am\ stars formation. 

If the donor in an \am\ type system was initially a white dwarf, there are no reasons to expect that abundances
in  the  matter  lost by it vary in the course of evolution. They must to correspond to the ``standard'' chemical 
composition of a stellar core 
that experienced hydrogen burning, with small variations due to the differences in the masses of 
precursors of white dwarfs. Similar abundances will have matter lost by He-stars that filled Roche lobes soon after TAMS.
However,  since initial  $ Y \approx 1 $, while initial  
$ X_{\rm  C} \sim 10^{-4} $,  $ X_{\rm  N} \sim 10^{-2} $,  $ X_{\rm  O} \sim 10^{-3} $, detection of
even   slight enrichment in C and O or 	impoverishment in N  may indicate possibility of formation of \am\ stars with
 He-star donors.   

\section{Conclusion}
\label{sec:concl} 

Above, we presented results of evolutionary computations for semidetached low-mass helium components in the systems with
white dwarf accretors. Evolution was considered as conservative in mass, but accompanied by angular momentum loss via
gravitational waves radiation. We may summarize our main results as follows.

Evolution of binaries under consideration only weakly depends on the mass of helium star, total mass of the system,
and evolutionary state of the donor at the instant of RLOF. The minimum  \porb = 9.3 -- 10.9\,min, the
masses of the donor at $\porb_{, min}$ are  0.20\,\ms -- 0.26\,\ms\ (with exception of initially very far-evolved system
$(0.65+0.8)$\,\ms, $\porb_{,0}=90$\,min.) which, in fact, represents a transitional case between systems with initially
nondegenerate and degenerate donors). In the period range $20 - 40  $ min, which may be adequately described by
our models and which hosts a significant fraction of all 
observed \am\ stars,  the scatter of \md\ for a given \porb\  does not exceed a factor $ \sim 2.5 $. 
 
Our computations are limited by $ \mhe \apgt 0.02$\,\ms. With further decrease of stellar mass thermal timescale of the donor 
becomes shorter than the angular momentum loss timescale, the stage of adiabatic donor expansion comes to the end,
and the matter of the donor becomes more degenerate as it cools down. Qualitative comparison with the computations for 
initially arbitrary degenerate helium white   dwarfs allows to suggest that at $\porb \apgt$ (40--45)\,min ``white-dwarf'' and
``helium-star'' families of \am\ stars merge and only analysis of the chemical composition of the matter lost by the donor
is able to discriminate between different scenarios of formation. Such an analysis is carried out in continuation of the present study. This result also shows that existing theoretical models of the population of \am\ stars might need certain revision, since they were carried out under assumption that helium stars evolve till Hubble time with the same $M-R$ relation.  

Initial periods of the most typical, according to current understanding, progenitors of \am\ stars with initially nondegenerate helium
donors are confined to about (20 -- 120)\, min. This implies a very narrow range of initial separations of components (for instance,
2\,\rs -- 7\,\rs\  for  $ \mhe + \mwd = 0.35\,\ms\ + 0.5\,\ms $ pair). This interval is defined by a combination of different parameters.
A binary passes through two common envelope stages. The ``standard'' equation for the variation of orbital separation of components
based on energy balance between binding energy of the mass-losing star and orbital energy of the system (Webbink, 1984; 
de Kool et al., 1987) is 
\begin{equation}
\frac{a_f}{a_i}= \frac{M_{1,c}}{M_1}\left[ 1+\left( \frac{2}{\alpha\lambda r_{1,L}}\right) 
\left( \frac{M_1-M_{1,c}}{M_2}\right) \right]^{-1}, 
\end{equation}  
where the indexes  $ i$ and $f $ label initial and final separation of components, 
$\alpha$ -- is the parameter of common envelope efficiency, $\lambda$ -- is the parameter of the binding energy of the 
stellar envelope,
$M_1  $ and $ M_{1,c} $ are initial mass of mass-losing star and the mass of its remnant, 
$   r_{1,L}$ is dimensionless radius of the star at the beginning of mass transfer,
$ M_2 $  is the mass of companion. 

For every common-envelope stage the values of $\alpha$\ and $\lambda$\ are, most probably, different and depend on  $a_i $
 at the beginning of the stage; the value of $a_f  $ after the second common envelope depends on ``initial-final'' mass relations 
for progenitors of white dwarf and helium star, which, in their own turn, depend on their evolutionary state at the instant of RLOF. 
This suggests that formation of \am\ stars needs a very fine ``tuning'' of evolutionary parameters. The treatment of stellar 
evolution incorporated in the population synthesis codes that predict existence of  \am\ stars with initially nondegenerate He-donors is,
at the moment, inevitably too crude for account of all subtleties of parameters. If \am\ stars with confidently established
presence of He-burning products signatures in their spectra will not be detected, this may mean that the ``necessary''
combination of parameters is not realized in the Nature or this combination is such, that all He-donors overflow Roche lobes prior
to ignition of He or when He burns very weakly. Such a possibility is suggested by positions of \am\ and V803~Cen in the
period-mass diagram (Fig.~  \ref{fig:pm}). Since initial abundance of He in stars is close to 1, while abundances of C, O, N are by 2 to 4 orders of magnitude lower, even  slight enhancements of $ X_{\rm  C} $ and  $ X_{\rm  O}  $ or reduction of 
$ X_{\rm  N}  $ may indicate possibility of formation of \am\ systems with He-star donors.

Above-discussed scenario of evolution needs some caveats. Evolution of every considered system has two phases -- prior and after 
period minimum. In the systems with  $ M_{2,0} \approx (0.35 -0.40)$ and $ \porb_{,0} \le 120$\,min.  which we consider as typical,
in the first stage of mass-exchange $ 2.5 \times 10^{-8} \aplt \md \aplt 10^{-7}$\,\myr. Livne (1990), Livne and Glasner (1991), 
Livne and Arnett (1995) and other authors have shown that for accretion rate  $\md \sim 10^{-8}$\myr\ accumulation of    $ \sim 0.1 $\,\ms\ of He at
the surface of CO white dwarf may result in detonation of He which may initiate detonation of carbon in the center of white dwarf 
and, presumably, SN~Ia\footnote{``Double-detonation'' or ``edge-lit detonation''
 (ELD).}.  Modeling by means of population synthesis has shown that ELDs may be dominating mechanism of SN~Ia in young
($\lesssim 1$\,Gyr) populations (see, e.g., Branch et al., 1995; Hurley et al., 2002; Yungelson, 2005b). However, models of light-curves
(Hoeflich  and Khokhlov, 1996; Hoeflich et al., 1996) and spectra (Nugent et al., 1997)  of objects experiencing ELDs revealed that  
they disagree with observations. But this conclusion is by no means final in the absence of more detailed calculations.
Double-detonation, if it occurs, may prevent transformation into \am\ stars of some ($ \sim 40\%$) of their potential
precursors (Nelemans et al., 2001). Note, for instance, that double-detonation may prevent formation of an \am\ star with initial parameters $ \mhe + \mwd =$ (0.65+0.80)\,\ms.
It is possible that instead of a detonation of a massive He-layer recurrent Nova-scale helium flashes  occur (see, e.g., Bildsten et al., 2007 and 
references therein). The flashes may be accompanied by mass and momentum loss from the system and change the behavior of evolutionary sequences. The problem of 
unstable He-burning at the surface of accreting white dwarfs and its possible impact on the evolution of \am\ stars needs additional investigation. 
 
\vskip 1mm
\textit{Availability of results}:
Detailed~results~of~ computations~may~be~found~at \\
\verb=www.inasan.ru/~lry/HELIUM_STARS/=.  

\vskip 3mm
{\small The author acknowledges G. Nelemans
for numerous discussions of the problems related to \am\ stars, thorough reading of the manuscript and useful comments. 
G. Roelofs and C. Deloye are acknowledged for useful exchange of opinions in the course of this study. 
The author is grateful to P.P. Eggleton for providing his evolutionary code and
to C. Deloye for providing the data of his calculations.
Present study was supported in part by RFBR grant  07-02-00454,
Russian Academy of Sciences program ``Origin and evolution of stars and galaxies'' and NSF grant    PHY05-51164.
A part of calculations for this study was carried out in the California NanoSystems Institute at UC Santa Barbara with support of NSF grant   CHE-0321368.
}

\section*{REFERENCES}
\label{sec:refs}

\begin{enumerate}
\item S.F. Anderson, D. Haggard, L. Homer, et al.,   \aj \textbf{130}, 2230 (2005). 
\item  S.F. Anderson, A.C. Becker, D. Haggard, et al., eprint arXiv:0802.2240 (2008).
\item L. Bildsten, K. Shen, N. Weinberg, et al., \apj \textbf{662}, L95 (2007).	
\item  D. Branch, M. Livio,  L.~R. Yungelson, et al., \pasp \textbf{107}, 1019 (1995). 
\item   G.~R. Caughlan and  W.~A. Fowler, Atomic Data and Nuclear Data Tables  \textbf{40}, 283 (1988). 
\item G.~R. Caughlan, W.~A. Fowler, M.~J. Harris, et al.,  Atomic Data and Nuclear Data Tables  \textbf{32}, 197 (1985).
\item M. de Kool, E.~P.~J. van den Heuvel, E. Pylyser \aap \textbf{183}, 47 (1987). 
\item C. Deloye, R.~E. Taam, C. Winisdoerffer, et al., \mnras \textbf{381}, 525 (2007). 
\item  P.~P.Eggleton, \mnras \textbf{151} 351 (1971).
\item  E.~V.  Ergma and A.~V. Fedorova,  Astrophys. Space Sci. \textbf{163}, 143 (1990).
\item    C.~R. Evans, I. Iben Jr., L. Smarr,  \apj  \textbf{323}, 129 (1987). 
\item J. Faulkner, \apj \textbf{70}, L99 (1971).
\item V. Gokhale, X.~M. Peng, J. Frank, \apj \textbf{655}, 1010 (2007).
\item D. Hils, P.~L. Bender, \apj \textbf{537}, 334 (2000). 
\item 	Z. Han, Ph. Podsiadlowski, P.~F. Maxted, et al., \mnras \textbf{336}, 449 (2002). 
\item   J.~R. Hurley,  C.~A. Tout,  O.~R. Pols,  \mnras \textbf{329}, 897 (2002).
\item P. Hoeflich and A. Khokhlov, \apj  \textbf{457}, 500 (1996). 
\item P. Hoeflich, A. Khokhlov, M.~M. Phillips, et al., \apj \textbf{472}, L81 (1996).
\item	 I. Iben Jr., \apj \textbf{353}, 215 (1990). 
\item	 I. Iben Jr., A.~V. Tutukov, \apj \textbf{315}, 727 (1987).
\item I. Iben Jr., A.~V. Tutukov, \apj \textbf{370}, 615 (1991).
\item R. Kippenhahn and A. Weigert, Zs. Astrophys. \textbf{65}, 251 (1967).
\item D.~L. Lambert and M.~H. Slovak, \pasp \textbf{93}, 477 (1981).
\item L.D. Landau and E.L. Lifshitz  \textit{Classical Theory of Fields}, 3d ed., Oxford:Pergamon (1971).
\item E. Livne, \apjl  \textbf{354}, L53 (1990).
\item E. Livne and D.  Arnett, \apjl  \textbf{452}, 62 (1995).
\item E. Livne and A. Glasner, \apj  \textbf{370}, 272 (1991).
\item T.~R. Marsh, K. Horne, S. Rosen, \apj  \textbf{366}, 535 (1991).
\item T.~R. Marsh, J.~H. Wood, K. Horne, et al., \mnras  \textbf{274}, 452  (1995).
\item  T.~R. Marsh, G. Nelemans, D. Steeghs, \mnras  \textbf{350}, 113 (2004). 
\item  T.~R. Marsh, V.~S. Dhillon, S.~P. Littlefair, et al., \textit{15th European Workshop on White Dwarfs, PASP Conf. Ser. 372, 2007} (Ed. R. Napiwotzki, M.~R. Burleigh),  p. 431. 
\item  P.~M. Motl, J. Frank, J.~E. Tohline, \apj  \textbf{670}, 1314 (2007).
\item  G. Nelemans, S.~F. Portegies Zwart, F. Verbunt, et al., \aap  \textbf{368}, 939 (2001).
\item   G. Nelemans and C.~A. Tout,
	\textit{White Dwarfs, NATO ASIB Proc. 105} (Ed.  D. de Martino, R. Silvotti, J.-E. Solheim, and R. Kalytis. Kluwer Acad. Publ. 2003), p. 359.
\item 	G. Nelemans, L. Yungelson, S.~F. Portegies Zwart, \mnras  \textbf{349}, 181 (2004).
\item P. Nugent,  E. Baron, D. Branch,  et al., \apj  \textbf{485}, 812 (1997). 
\item	 B. Paczy\'{n}ski, \acta \textbf{17}, 287 (1967a).
\item B. Paczy\'{n}ski, \acta \textbf{17}, 355 (1967b).
\item  B. Paczy\'{n}ski, \acta \textbf{21}, 1 (1971).
\item 	B. Paczy\'{n}ski, \acta \textbf{31}, 1 (1981).
\item 	Ph. Podsiadlowski, Z. Han, S. Rappaport, \mnras  \textbf{340}, 1214 (2003).	
\item   O.~R. Pols, C.~A. Tout, P.~P. Eggleton, et al., \mnras  \textbf{274}, 964 (1995).
\item 	K.~A. Postnov and L.~R. Yungelson, 	Living Reviews in Relativity, \textbf{9}, no. 6 (2006).
\item G.~H.~A. Roelofs, P.~J. Groot, G. Nelemans, et al., \mnras  \textbf{371}, 1231 (2006).	
\item  G.~H.~A. Roelofs, P.~J. Groot, G.~F. Benedict, et al., \apj  \textbf{666}, 1174 (2007a).	
\item  G.~H.~A. Roelofs, P.~J. Groot, G. Nelemans, et al., \mnras \textbf{379}, 176 (2007b).
\item 	G.~H.~A. Roelofs, G. Nelemans, P.~J. Groot, \mnras \textbf{382}, 685 (2007c).
\item  G.~J. Savonije,  M. de Kool, E.~P.~J. van den Heuvel, \aap \textbf{155}, 51 (1986).
\item J.-E. Solheim and  L.~R. Yungelson,
\textit{14th European Workshop on White Dwarfs, PASP Conf. Ser. 334} (Ed. D. Koester, S. Moehler), p. 387  (2005).
\item A. Stroeer and  A. Vecchio, 	Class. Quant. Grav. \textbf{23}, S809 (2006).
\item T.~E Strohmayer, \apj \textbf{608}, L53 (2004).
\item 	A.V. Tutukov and A.V. Fedorova, SvA,  \textbf{33}, 606 (1989).
\item	  A.V. Tutukov, A.V. Fedorova, E.V. Ergma., et al.,
SvAL, \textbf{11}, 52 (1985).
\item	 A.~V. Tutukov and L.~R. Yungelson, \acta \textbf{29}, 665 (1979).
\item  A.~V. Tutukov and L.~R. Yungelson,  Nauchnye Informatsii \textbf{49}, 3
(1981). 
\item 	A.~V. Tutukov and L.~R. Yungelson,  \mnras \textbf{280}, 1035 (1996).
\item R.~F. Webbink, \apj \textbf{277}, 355 (1984). 
\item L.~R. Yungelson, \textit{Interacting binaries: Accretion, Evolution, and Outcomes, AIP Conf. Proc. 797} (Ed. L. Burderi, L.~A.  Antonelli, F. D'Antona et al.), p. 1  (2005a).
\item  L.~R. Yungelson,  
\textit{White dwarfs: cosmological and galactic probes, ASSL 332} (Ed. E.~M. Sion, S. Vennes, H.~L. Shipman),  p. 163 (2005b).
\end{enumerate}

\end{document}